\PassOptionsToPackage{unicode}{hyperref}
\PassOptionsToPackage{hyphens}{url}
\documentclass[11pt, ]{article}

\usepackage{mathtools}
\usepackage{amsmath}
\usepackage{amsthm}
\usepackage{amssymb}
\usepackage[italicdiff]{physics}
\mathtoolsset{showonlyrefs}

\usepackage{iftex}
\ifPDFTeX
  \usepackage[OT1,T1]{fontenc}
  \usepackage[utf8]{inputenc}
  \usepackage{textcomp} % provide euro and other symbols
  \usepackage{ebgaramond}
  \usepackage{biolinum}
    \usepackage[scale=0.95,varl]{inconsolata}
\else % if luatex or xetex
  \usepackage[scale=0.95,varl]{inconsolata}
  \usepackage{newpxtext}
  \usepackage{mathpazo}
    \usepackage[scale=0.95]{biolinum}
  \fi
\ifLuaTeX
  \usepackage{selnolig}  % disable illegal ligatures
\fi
\IfFileExists{microtype.sty}{% use microtype if available
  \usepackage[]{microtype}
  \UseMicrotypeSet[protrusion]{basicmath} % disable protrusion for tt fonts
}{}

\allowdisplaybreaks
\usepackage[dvipsnames,svgnames,x11names]{xcolor}
\usepackage[margin=1in]{geometry}
\usepackage[format=plain,
  labelfont={bf,sf,small,singlespacing},
  textfont={sf,small,singlespacing},
  justification=justified,
  margin=0.25in]{caption}

\usepackage{sectsty}
\usepackage[compact]{titlesec}
\makeatletter
\newcommand\@shorttitle{}
\newcommand\shorttitle[1]{\renewcommand\@shorttitle{#1}}
\usepackage{fancyhdr}
\fancyheadoffset{0pt}
\makeatother
\renewenvironment{abstract}{
  \centerline
  {\large\sffamily\bfseries Abstract}\vspace{-1em}
  \begin{quote}\small
}{
  \end{quote}
}

\providecommand{\tightlist}{%
  \setlength{\itemsep}{0pt}\setlength{\parskip}{0pt}}\usepackage{longtable,booktabs,array}
\usepackage{calc} % for calculating minipage widths
\usepackage{etoolbox}
\makeatletter
\patchcmd\longtable{\par}{\if@noskipsec\mbox{}\fi\par}{}{}
\makeatother
\IfFileExists{footnotehyper.sty}{\usepackage{footnotehyper}}{\usepackage{footnote}}
\makesavenoteenv{longtable}
\usepackage{graphicx}
\makeatletter
\def\maxwidth{\ifdim\Gin@nat@width>\linewidth\linewidth\else\Gin@nat@width\fi}
\def\maxheight{\ifdim\Gin@nat@height>\textheight\textheight\else\Gin@nat@height\fi}
\makeatother
\setkeys{Gin}{width=\maxwidth,height=\maxheight,keepaspectratio}
\makeatletter
\def\fps@figure{htbp}
\makeatother
\newcommand{\aian}{\textsc{ai/an}}
\makeatletter
\@ifpackageloaded{caption}{}{\usepackage{caption}}
\AtBeginDocument{%
\ifdefined\contentsname
  \renewcommand*\contentsname{Table of contents}
\else
  \newcommand\contentsname{Table of contents}
\fi
\ifdefined\listfigurename
  \renewcommand*\listfigurename{List of Figures}
\else
  \newcommand\listfigurename{List of Figures}
\fi
\ifdefined\listtablename
  \renewcommand*\listtablename{List of Tables}
\else
  \newcommand\listtablename{List of Tables}
\fi
\ifdefined\figurename
  \renewcommand*\figurename{Figure}
\else
  \newcommand\figurename{Figure}
\fi
\ifdefined\tablename
  \renewcommand*\tablename{Table}
\else
  \newcommand\tablename{Table}
\fi
}
\@ifpackageloaded{float}{}{\usepackage{float}}
\floatstyle{ruled}
\@ifundefined{c@chapter}{\newfloat{codelisting}{h}{lop}}{\newfloat{codelisting}{h}{lop}[chapter]}
\floatname{codelisting}{Listing}

\makeatother
\makeatletter
\@ifpackageloaded{caption}{}{\usepackage{caption}}
\@ifpackageloaded{subcaption}{}{\usepackage{subcaption}}
\makeatother
\makeatletter
\@ifpackageloaded{tcolorbox}{}{\usepackage[skins,breakable]{tcolorbox}}
\makeatother
\makeatletter
\@ifundefined{shadecolor}{\definecolor{shadecolor}{rgb}{.97, .97, .97}}
\makeatother
\usepackage[]{natbib}
\newenvironment{CSLReferences}[2]{
\bibliography{references.bib}
\clearpage
}{}

\usepackage{hyperref}
\usepackage{url}
\hypersetup{
  pdftitle={Making Differential Privacy Work for Census Data Users},
  pdfauthor={Cory McCartan; Tyler Simko; Kosuke Imai},
  pdfkeywords={census data, differential privacy, open science},
  colorlinks=true,
  linkcolor={black},
  filecolor={Maroon},
  citecolor={VioletRed4},
  urlcolor={DodgerBlue4},
  pdfcreator={LaTeX via pandoc}}

\title{\sffamily\bfseries\huge\parfillskip=0pt
\rightskip=0pt plus .5\textwidth
\leftskip=0pt plus .5\textwidth
\emergencystretch=.3\textwidth Making Differential Privacy Work for
Census Data Users}
\shorttitle{Making Differential Privacy Work for Census Data Users}
\author{\textbf{Cory McCartan}
\\Center for Data Science\\New York University
\vspace{0.05in}
 \and \textbf{Tyler Simko}
\\Department of Government\\Harvard University
\vspace{0.05in}
 \and \textbf{Kosuke Imai}\footnote{
To whom correspondence should be addressed.
Email: \texttt{\href{mailto:imai@harvard.edu}{imai@harvard.edu}}.
Website: \url{https://imai.fas.harvard.edu/}.
Address:
1737 Cambridge Street, Cambridge, MA 02138.
The authors thank Christopher T. Kenny and Shiro Kuriwaki for helpful
comments and suggestions, and Michael B. Hawes of the Census Bureau's
Research and Methodology Directorate for sharing information about
various details of the Noisy Measurement File.}
\\Department of Government\\
Department of Statistics\\Harvard University
\vspace{0.05in}
 }
\date{October 7, 2023}

\begin{document}
\allsectionsfont{\sffamily}

\maketitle

\begin{abstract}
The U.S. Census Bureau collects and publishes detailed demographic data
about Americans which are heavily used by researchers and policymakers.
The Bureau has recently adopted the framework of differential privacy in
an effort to improve confidentiality of individual census responses. A
key output of this privacy protection system is the Noisy Measurement
File (NMF), which is produced by adding random noise to tabulated
statistics. The NMF is critical to understanding any errors introduced
in the data, and performing valid statistical inference on published
census data. Unfortunately, the current release format of the NMF is
difficult to access and work with. We describe the process we use to
transform the NMF into a usable format, and provide recommendations to
the Bureau for how to release future versions of the NMF. These changes
are essential for ensuring transparency of privacy measures and
reproducibility of scientific research built on census data.
\end{abstract}

\textbf{\textit{Keywords}}\quad census data~\textbullet~differential
privacy~\textbullet~open science
\ifdefined\Shaded\renewenvironment{Shaded}{\begin{tcolorbox}[frame hidden, enhanced, interior hidden, boxrule=0pt, sharp corners, borderline west={3pt}{0pt}{shadecolor}, breakable]}{\end{tcolorbox}}\fi

% USER BODY %%%%%%%%%%%%%%%%%%%%%%%%%%%%%%%%%%%%%%%%%%%%%%%%%%%%%%%%%%%%%%%%%%%%

\hypertarget{sec-intro}{%
\section{Introduction}\label{sec-intro}}

For the 2020 decennial census, the Census Bureau adopted a new
Disclosure Avoidance System (DAS) based on differential privacy. The DAS
was designed to protect the confidentiality of responses by injecting
statistical noise into the tabulations of a confidential individual
census dataset. The Bureau post-processed the resulting Noisy
Measurement File (NMF) to produce the final tabulated statistics.
Post-processing ensured that the final tabular statistics meet various
requirements, including non-negative counts, invariant state
populations, and consistency across different geographical levels of
census hierarchy.

However, as some of the Bureau's staff themselves warned,
post-processing also introduces biases in the tabulated statistics
{[}\citet{NAP2020}; see e.g., page 42{]}.\footnote{See also the
  transcript of a Bureau webinar:
  \url{https://www2.census.gov/about/training-workshops/2021/2021-05-13-das-transcript.pdf}.}
Some of these biases, like inflating the average population of some
zero-population areas, were expected, others were discovered by
concerned data users and in cases required changes to the DAS
\citep{dwork2021letter, kenny2021impact, ncai2021, jason2022, kenny2023comment, scariano2022balancing}.

The NMF, which is an intermediate product of the DAS, is an invaluable
resource for researchers and other Census data users to understand the
error introduced by the DAS. The NMF is also essential for performing
statistically valid analyses of Census data that properly account for
DAS-introduced error \citep{kenny2023evaluating}.

In 2021, however, the Bureau did not make the NMF available and only
released the final (post-processed) tabulated statistics. Through an
open letter \citep{dwork2021letter}, a Freedom of Information Act
request, and subsequent litigation \citep{nmflawsuit}, dozens of
scholars and data users urged the Bureau to release the NMF, which
allows one to understand and correct potential biases induced by the
post-processing procedure. Following these developments, in April 2023,
the Bureau released a demonstration NMF based on the 2010 Census data,
and subsequently in June made available the NMF for the 2020 P.L.94-171
redistricting data.\footnote{Prior to these developments, a member of
  the DAS team mentioned the possibility of releasing the NMF \citep[see
  e.g., Section 11.2]{NAP2020}.} Furthermore, in May 2023, the Bureau
also published the 2020 Demographic and Housing Characteristics File
(DHC) and plans to publish its NMF later this year.

We commend the Bureau's decision to provide the NMF, which will help
advance social science research, improve policy decisions, and further
strengthen the DAS itself. To maximize the benefits of the released NMF,
however, we believe that the Bureau must substantially improve the way
in which the NMF is formatted and released \citep{mccartan2023}. Below,
we first explain several obstacles many end users are likely to face
when accessing, processing, and using the released NMF for statistical
analyses. We then provide recommendations for how the Census Bureau
could release the NMF going forward in a manner and format that would
address these issues.

\hypertarget{sec-nmf}{%
\section{The Census Noisy Measurement File}\label{sec-nmf}}

\begin{figure}

{\centering \includegraphics[width=\textwidth,height=2.5in]{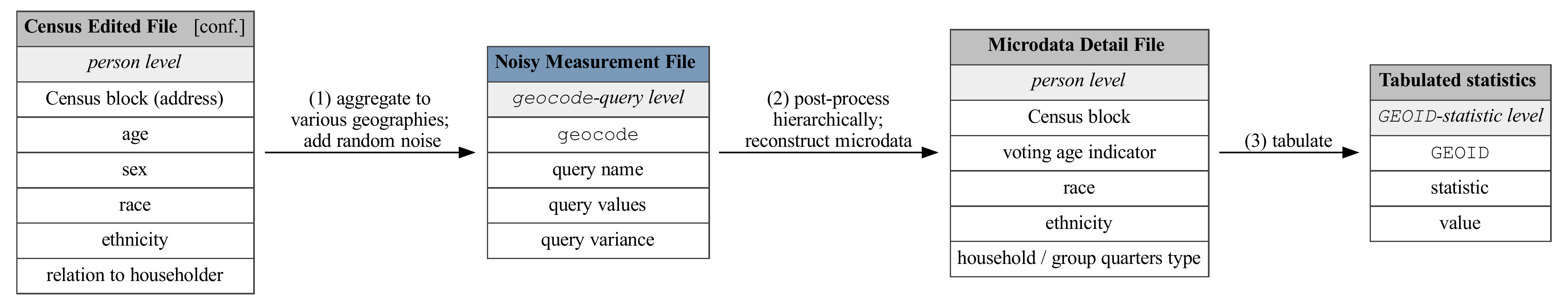}

}

\caption{\label{fig-overview}How the Census Bureau produces
differentially private census tabulations for person-level redistricting
(P.L. 94-171) data, starting from the confidential Census Edited File.
Step (2) is the core of the TopDown Algorithm, detailed in
\citet{topdown}. Together, steps (1) and (2) comprise the Disclosure
Avoidance System (DAS).}

\end{figure}

After collecting census questionnaires from the U.S. population, the
Census Bureau fills in missing responses using other data sources and
imputation methods. The resulting confidential dataset is called the
Census Edited File (CEF), which has a row for every person in the U.S.
To convert the CEF into statistics that can be reported publicly, the
DAS proceeds in two steps, as illustrated in Figure~\ref{fig-overview}.
The process described in this section is a simplification of the full
workings of the DAS; further details may be found in \citet{topdown}.

First, a series of \emph{queries} are made for every geographic unit on
the NMF \emph{geographic spine}: states, counties, tracts, optimized
block groups, and blocks.\footnote{The geographic spine is used for the
  DHC file is different from the one used for the P.L.94-171
  redistricting data.} Notably, the geographic spine is split between
regions inside and outside of American Indian / Alaska Native (\aian)
areas . This means that, for example, a county containing a Native
reservation is split into two geographic units: the \aian{} area and the
remainder of the county. Each of these has queries made separately.
These queries are contingency tables for various statistics. One example
is the ``total population'' query which counts the total number of
people in the geographic unit. Another is the ``voting age by Hispanic''
query which reports a 2-by-2 table of counts of people in those binary
categories.

For the decennial census redistricting data, there are 11 total queries
for each geography, including a ``detailed'' query which is the full
contingency table of voting age, Hispanic status, race, and household
type, with 2,016 values. The counts that make up each of these queries
have a certain amount of discrete Gaussian noise added by the DAS
\citep{canonne2020discrete}, in accordance with a privacy loss schedule
set by the Bureau, which is recorded in the \texttt{variance} column of
the NMF. It is this step that ensures the released statistics meet a
certain standard of differential privacy
\citep{bun2016concentrated, kifer2022bayesian}.

Second, the set of noised queries across all geographic units and levels
is post-processed using a multi-pass optimization routine. The goal of
this post-processing step is to produce a set of counts that is
self-consistent across geographic units and different statistics, and
respects commonsense constraints like having nonnegative population
counts. From these consistent counts, synthetic person-level data can be
reconstructed, yielding the Microdata Detail File. Unlike the CEF, the
exact format of the Microdata Detail File depends on the particular
census product. It is then straightforward to tabulate this microdata to
produce traditional census statistics for each geography in the country.

\hypertarget{sec-proc}{%
\section{Processing the Noisy Measurement File}\label{sec-proc}}

We identified four key obstacles that make it difficult to directly use
the 2020 NMF released in June 2023 \citep{NMF2023}:

\begin{enumerate}
\def\labelenumi{\arabic{enumi}.}
\tightlist
\item
  Collecting the documentation necessary to open and parse the NMF;
\item
  Converting the NMF data files to a familiar format amenable to
  computer and statistical analysis;
\item
  Navigating the many-to-many mapping between NMF \emph{queries} and
  tabulated \emph{statistics}; and
\item
  Connecting the \emph{geocodes} used in the NMF to traditional census
  \emph{tabulation geographies}.
\end{enumerate}

This section describes these obstacles and our process for overcoming
them.

The NMF was accompanied by sparse documentation which was not alone
sufficient to reconstruct the format and coding of the data. The Bureau
has made the code that implements the DAS publicly available, but this
code as well is sparsely documented and highly unlikely to be accessible
to non-experts. Particularly troubling is that the codebook for the
variables (Male=0, Female=1, etc.) exists only as a set of undocumented
Python files in a folder in this code repository. This problem is
compounded for statistics that involve cross-tabulations of base
variables. For example, a ``sex by Hispanic'' table could be reported in
the order of (Non-hispanic male, Non-hispanic female, Hispanic male,
Hispanic female), or it could be reported in the order of (Non-Hispanic
male, Hispanic male, Non-Hispanic female, Hispanic female). There is no
clear documentation as to which of these orders variables are combined
in, and the actual order depends on the way the NumPy array library for
Python works. Despite these challenges, by manually copying the codebook
from the Python code and running numerous experiments on the NMF data,
we were able to reconstruct the structure and variable coding of the NMF
ourselves.\footnote{The code is available at
  \url{https://github.com/uscensusbureau/DAS_2020_Redistricting_Production_Code}}

\begin{figure}

{\centering \includegraphics[width=1\textwidth,height=\textheight]{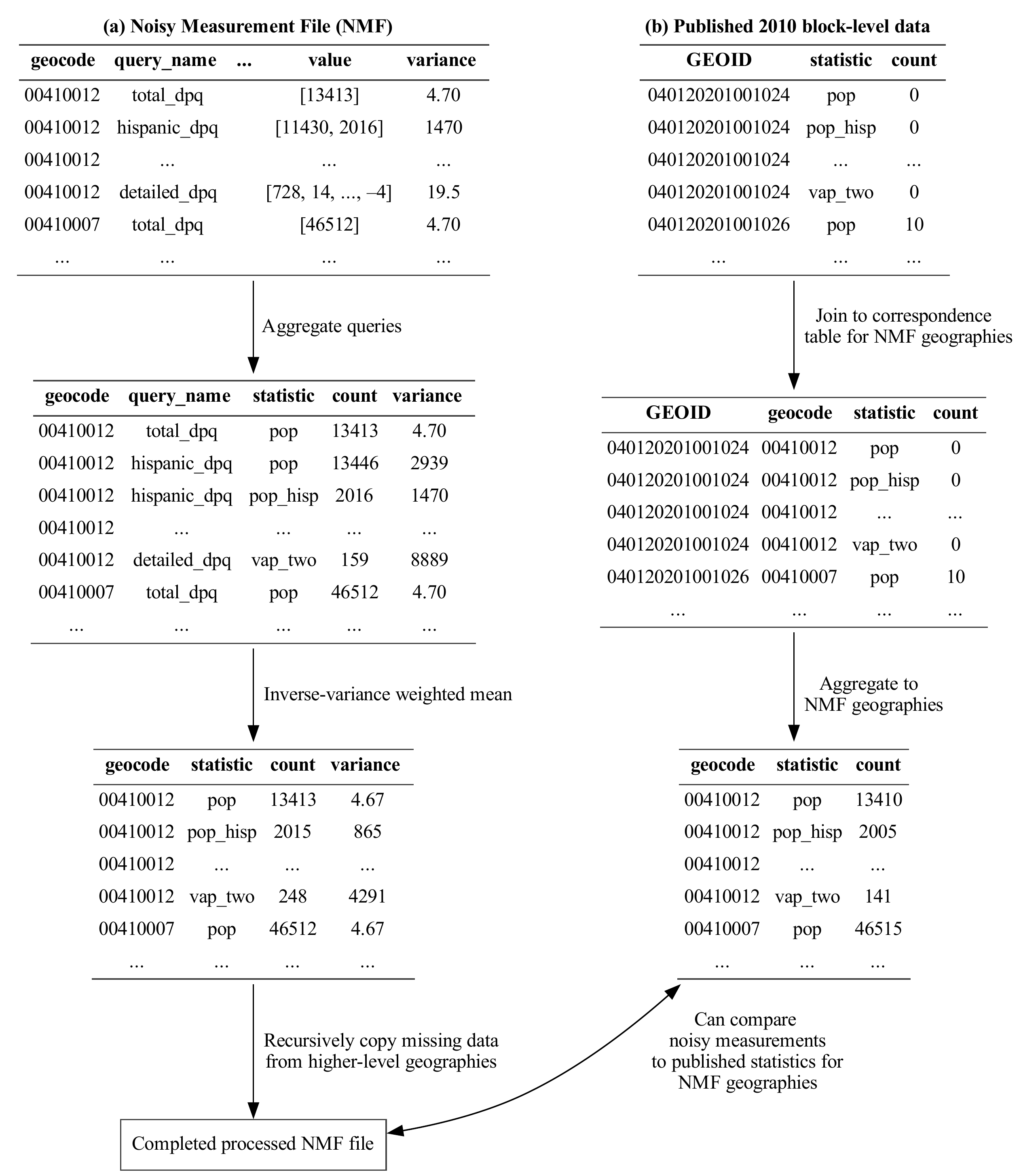}

}

\caption{\label{fig-workflow}\textbf{(a)} Our workflow for processing
the Noisy Measurement File into a usable form, illustrated as samples
from the NMF data table as it is transformed, and \textbf{(b)} the
necessary workflow for processing published statistics so that they are
directly comparable with noisy measurements. The \texttt{\_dpq} suffix
is used by the Bureau to indicate a differentially private query (as
opposed to a traditional tabulation statistic).}

\end{figure}

The NMF is made available as a series of large Apache Parquet files, a
flexible archival storage format that underpins much of modern big data
infrastructure. This flexibility allows the Bureau to store the NMF in a
non-rectangular format: each row of the NMF contains normal records like
the query name and variance, but also a variable-length vector
containing the query results (collapsed from a multidimensional
contingency table). This nested data structure unfortunately cannot be
directly analyzed using standard tools designed for rectangular data.

As described in Section~\ref{sec-nmf}, at this stage each row in the NMF
corresponds to a single query run for a particular geographic unit (see
the top panel of Figure~\ref{fig-workflow}(a)). Each query's result can
be aggregated to produce an estimate of various traditional census
statistics. For example, the ``voting age by Hispanic'' query result for
some geography \(g\) is stored in a single row of the NMF as the
following vector with 4 elements: \[
    m_g = (\overbrace{1050}^{\mathclap{\text{Non-Hisp. children}}},\ 
    \underbrace{204}_{\mathclap{\text{Hisp. children}}},\ 
    \overbrace{10050}^{\mathrlap{\text{Non-Hisp. adults}}},\ 
    \underbrace{1812}_{\mathclap{\text{Hisp. adults}}})^\top.
\]

This vector, which represents a \(2\times 2\) table, can be summed along
each dimension to produce (noisy) estimates of the voting age population
or the Hispanic population, or all its entries can be summed to estimate
the total population of the geographic unit.

Thus, each traditional tabulated statistic, like total or voting-age
population, has multiple noisy ``versions'' spread across multiple
queries. For example, the Hispanic population can be estimated not just
by the ``Hispanic'' query, but by the ``voting age by Hispanic'' query,
the ``race by Hispanic'' query, the ``voting age by race by Hispanic''
query, and the detailed query. For further analysis, data users must map
the queries to statistics of interest, possibly combining information
across multiple queries.

To address this challenge, we developed a series of aggregation matrices
which, when multiplied by a vector of noisy query values, produces a set
of tabulated statistics. For example, the aggregation matrix for the
``voting age by Hispanic'' query shown above would be applied as
follows: \[
\overbrace{A}^{\mathclap{\text{Agg. matrix}}}
\underbrace{m_g}_{\mathclap{\text{Query result}}}
= \begin{pmatrix}
1 & 1 & 1 & 1\\
0 & 0 & 1 & 1\\
0 & 1 & 0 & 1\\
0 & 0 & 0 & 1\\
\end{pmatrix} 
\begin{pmatrix}1050\\204\\10050\\1812\end{pmatrix} 
= (\overbrace{13116}^{\mathclap{\text{Pop. estimate}}},\ 
    \underbrace{11862}_{\mathclap{\text{VAP estimate}}},\ 
    \overbrace{2016}^{\mathclap{\text{Hispanic estimate}}},\ 
    \underbrace{1812}_{\mathclap{\text{Hisp. VAP estimate}}})^\top.
\]

Applying these aggregation matrices to each of the queries produces
multiple noisy estimates of each of the desired tabulated statistics.
Applying them again to the variance of each of the queries (which is
included in the NMF) calculates the variance of each of these estimates,
since each entry in \(m_g\) is independent of the others. The output of
this step is shown in the second panel of Figure~\ref{fig-workflow}(a).
Then, to integrate these estimates, we take an inverse-variance weighted
mean, which produces a minimum-variance estimator of each (true)
tabulated statistic.\footnote{This induces some correlation between
  noisy measurements \emph{within} each geography, but maintains
  independence between geographies and across geographic levels. Where
  this correlation is statistically unhelpful, aggregation matrices can
  be built that only take the minimum-variance query for each statistic,
  sacrificing lower variance for ensuring completely independent
  statistics.} The output of this step is shown in the third panel of
Figure~\ref{fig-workflow}(a).

Mapping queries to statistics yields data that look similar to the
tabulated tables traditionally released by the Bureau: each row
corresponds to a tabulated statistic for a particular geographic unit
(along with the error variance). But the similarity is only
superficial---the geographic units in the NMF, identified by numeric
\emph{geocodes}, do not directly correspond to traditional census
tabulation geographies, identified by so-called \emph{GEOIDs}. Each
Census block with non-zero population has a unique associated geocode.
Geocodes are combinations of traditional census information contained in
\emph{GEOIDs} (e.g.~FIPS codes) and information on TopDown processing
steps (e.g.~an indicator for whether the geography is on the \aian{}
spine or not). A full explanation is provided in \citet[Table
1]{kenny2023evaluating}, and original detail is available in the NMF
documentation \citep[see, e.g., footnotes 14-15]{nmfdocumentation}.

While blocks correspond to blocks, every other geography aside from the
country as a whole is different \citep{cumings2022geographic}. Part of
this difference is due to separating the geographic spine for regions
inside and outside of \aian{} areas as mentioned above. This decision
means that states, counties, tracts, and optimized block groups are
split between \aian{} regions and must be re-combined for many analyses.
Traditional block groups are also replaced with ``optimized block
groups,'' which are designed to minimize DAS-induced error for specific
off-spine geographies such as Census places. Finally, any geography with
no housing units is removed from the NMF.\footnote{Formally, the Bureau
  imposes a \emph{structural zero} requirement that blocks with no
  housing units report zero population.}

An additional complication is that when a geographic unit contains only
a single subunit (for example, a tract containing only one block group),
the subunit's privacy loss budget is reallocated to the higher-level
unit, and the subunit is removed from the NMF \citep[see][]{topdown}.
This creates ``holes'' in the NMF and substantially increases the
difficulty of analyzing the NMF.

Thankfully, a full listing of (populated) block geocodes exists in
auxiliary constraint files released alongside the NMF. Since the NMF
block-level geocodes can be split into two pieces, with the second piece
corresponding to the block's tabulation GEOID, this allows us to build a
correspondence table that links every census block to its traditional
tabulation geographies and NMF-specific geographies. Finally, by
recursing down the geographic hierarchy and identifying geographies
which are missing in the NMF but present in the auxiliary data, we can
copy data from higher geographic levels and fill the ``holes'' in the
NMF.

The correspondence tables linking NMF blocks to traditional tabulation
geographies are only partially satisfying. They allow analysts to
aggregate any block-level data, such as the final published census data
(the right table in Figure~\ref{fig-overview}) to both NMF and
traditional geographies. Yet, because noisy measurements are not
consistent across geographic levels---the sum of tract populations does
not equal the county population, in general---there are many ways to
aggregate NMF data to larger geographies. The simplest way---aggregating
NMF block-level data---also produces far noisier estimates, which may be
practically unusable for many purposes.

A better way involves aggregating noisy estimates from the largest
possible NMF geographies. For example, to produce an NMF-based estimate
of a school district's population, we would add up the NMF estimates of
all the tracts contained fully within the district, then all the
optimized block groups that are contained fully within the remainder,
and then finally the remaining blocks. If applied to all school
districts in the country, this process will produce a minimum-variance
unbiased estimates of the true tabulated populations, under the
constraint that each of the district estimates remain statistically
independent of the others and are built out of only geographies
contained in each district. This approach is used in
\citet{kenny2023evaluating}. Further research is warranted, however, in
establishing minimum-variance estimators under other constraints, and
evaluating the tradeoffs between different estimation approaches.

For now, we recommend that most researchers interested in comparing NMF
data to published tabulations should proceed by aggregating published
block data to NMF geographies (rather than the other way around). This
is illustrated in Figure~\ref{fig-workflow}(b).

\hypertarget{sec-rec}{%
\section{Recommendations for the Census Bureau}\label{sec-rec}}

Based on our experience obtaining the NMF and building tools to load,
format, and analyze it, we offer the following recommendations that will
significantly improve the usability of the NMF. Our recommendations are
focused on facilitating the main use cases of the NMF: understanding the
DAS-introduced error in published statistics and performing
statistically valid inference on those data.

\begin{enumerate}
\def\labelenumi{\arabic{enumi}.}
\item
  Data access and documentation

  \begin{enumerate}
  \def\labelenumii{(\alph{enumii})}
  \item
    Host the raw NMF files on an easy-to-use publicly-accessible
    website. The NMFs were first posted to the Globus service, which
    required users to create an account and install their third-party
    software locally in order to download any data, even the
    documentation. Some of the NMFs have been released on other
    platforms, such as the Harvard Dataverse; others have not. Ideally
    users could easily (and programmatically) download the individual
    Parquet files that make up the NMF from the Bureaus's own website.
  \item
    Consider un-nesting the NMF file before publication so that the data
    are in a rectangular format, with each query's histogram bins fully
    labeled. In addition to making the data much more amenable to
    standard data processing tools, the inclusion of labels would
    additionally obviate the need for users to manually copy values from
    codebooks.
  \item
    In the event that further reformatting and aggregation is performed
    by the Bureau on the NMF (see below for specific recommendations),
    provide API access to these reformated data, similar to the API for
    existing decennial census tabulations or ACS results.
  \item
    Produce centralized NMF documentation and make it available
    separately from the data. This documentation should explain the
    high-level structure of the NMF and its relation to published
    decennial statistics and tabulation geographies. The documentation
    should also contain sufficient detail and code books needed to
    properly read and format the file. The existing documentation
    provided alongside the raw NMF is around 8 pages long and, while
    technical, does not contain enough detail on its own to allow
    researchers to work with the NMF.
  \end{enumerate}
\item
  Data parsing and formatting

  \begin{enumerate}
  \def\labelenumii{(\alph{enumii})}
  \item
    Fill in NMF geographies missing due to privacy budget reallocation
    with data copied from higher-level geographies. While the original
    NMF reflects the structure of the geographic spine used as input to
    the DAS, without this filling-in, researchers are unable to use the
    NMF to evaluate bias and noise or make statistically valid
    inferences from privacy-protected data.
  \item
    In the event that the un-nested NMF is too large for users to work
    with, release code in standard data analysis languages (R, Python,
    Stata, etc.) that demonstrates how to load the NMF in and attach
    labels to the noisy queries.
  \end{enumerate}
\item
  Connecting NMF queries to tabulated statistics

  \begin{enumerate}
  \def\labelenumii{(\alph{enumii})}
  \item
    Provide aggregation specifications that link queries to traditional
    tabulation statistics in the redistricting or DHC files. These
    specifications should be designed as tables that can be joined to
    the NMF and then summarized to produce the necessary tabulations.
  \item
    Produce an additional version of the NMF for which these
    aggregations have already performed, and the results possibly
    combined using inverse-variance weighted means, so that each
    statistic has a single minimum-variance estimate. As described in
    Section~\ref{sec-proc}, this aggregation and weighting can be done
    independently across geographies, preserving the key statistical
    advantage of the NMF data---independent, unbiased noise---while
    making it directly usable for downstream tasks.
    Aggregation-and-weighting also significantly reduces the size of the
    data, allowing the output to be hosted by the Bureau and processed
    more easily by most end users. Other than the differences in the
    geographic spine, this data product could look much like the
    American Community Survey tabulations, which are accompanied by a
    margin of error estimate.
  \end{enumerate}
\item
  Connecting the geocodes used in the NMF to traditional census
  tabulation geographies

  \begin{enumerate}
  \def\labelenumii{(\alph{enumii})}
  \item
    Provide full block assignment files (BAFs) for NMF geographies,
    including blocks with no housing units.
  \item
    Produce shapefiles, for the NMF geographies so that they can be used
    directly in analyses, if desired, and compared to geographies which
    are of interest to end users, such as school or voting districts.
  \item
    Provide generalized geography assignment files which describe the
    best way to build each traditional tabulation geography (on- and
    off-spine) out of NMF geographies, following the approach discussed
    in Section~\ref{sec-proc}. Such generalized assignment files would
    enable researchers to have independent, unbiased noisy measurements
    of statistics of interest for their desired geographies, which are
    also minimum-variance.\footnote{Specifically, this would ensure that
      resulting estimates are minimum-variance, subject to the
      constraint that they are built only out of NMF geographies fully
      contained within tabulation geographies. This in turn means that
      estimates be statistically independent across geographies.}
  \end{enumerate}
\end{enumerate}

If implemented, these recommendations would produce a NMF that is much
more accessible for and familiar to data users, enabling them to focus
on developing methods to analyze noisy census measurements, and
answering substantive research questions.

\hypertarget{sec-conc}{%
\section{Conclusion}\label{sec-conc}}

We commend the Census Bureau for their decision to release the Noisy
Measurement File (NMF), which has the potential to allow Census data
users to understand and account for the DAS-introduced error for a wide
array of analyses. In addition, the NMF demonstration data release
offered a potential opportunity to discuss improvements to data
usability \emph{before} a wide range of Census stakeholders begin to
build new data pipelines to incorporate the NMF into their
work.\footnote{The Census Bureau has announced multiple 2020 NMF data
  releases.
  \url{https://www.census.gov/newsroom/press-releases/2023/2020-census-data-products-schedule-updates.html}}
Unfortunately, little change was made in the final release of the NMF to
overcome the obstacles we identified in the initial version of this
paper.\footnote{Some positive changes were made, including publishing
  copies of some of the NMFs on the Harvard Dataverse and Amazon Web
  Services Open Data Registry, and publishing copies of the READMEs that
  accompany the NMFs on the Census Bureau's FTP server.}

The NMF, as currently released, is difficult to process and use for the
kinds of tasks that will allow practitioners to evaluate and correct for
DAS-introduced error in their analyses. Fortunately, many of these
difficulties are logistical and not due to scientific necessity. We
believe that adopting our recommendations will result in a more
accessible and useful NMF for many Census data users. The Census Bureau,
as the developer of the DAS and of course the principal expert on Census
data, is best-positioned to make these changes and produce a data
product that is both statistically sound and practically useful. Our
recommendations are consistent with the Bureau's own stated goals around
data usability, such as their pioneering Enterprise Data Lake (EDL)
initiative to modernize data collection, processing, and
dissemination.\footnote{\url{https://www.census.gov/newsroom/blogs/research-matters/2022/10/technology-transformation.html}}

Data releases from the US Census Bureau serve as the backbone for a
great deal of scientific analyses and policy decisions. We hope these
discussions around data usability, accuracy, and privacy in the Census
can continue to serve as a useful precedent for data providers and data
users in the future.

\hypertarget{refs}{}

\begin{CSLReferences}{0}{0}\end{CSLReferences}

% END BODY %%%%%%%%%%%%%%%%%%%%%%%%%%%%%%%%%%%%%%%%%%%%%%%%%%%%%%%%%%%%%%%%%%%%%

\end{document}